\documentclass[aps,prc,twocolumn,showpacs,dvips]{revtex4-1}
\usepackage{graphicx}

\usepackage[tbtags]{amsmath}
\usepackage{bm,amssymb}
\usepackage{mathptm}
\newcommand{\br}{{\bm{r}}}
\newcommand{\bp}{{\bm{p}}}

\renewcommand{\Im}{\mathop{\rm Im}}

\newcommand{\w}{w}

\newcommand{\etal}{\textit{et al.}} 

\begin{document}

\title{Nascent fragment shell effects on the nuclear fission processes \\
in semiclassical periodic orbit theory}

\author{Ken-ichiro Arita,$^1$ Takatoshi Ichikawa$^2$ and
Kenichi Matsuyanagi$^{2,3}$}
\affiliation{
$^1$Department of Physics, Nagoya Institute of Technology,
Nagoya 466-8555, Japan \\
$^2$Yukawa Institute for Theoretical Physics,
Kyoto 606-8502, Japan \\
$^3$Nishina Center, RIKEN, Wako 351-0198, Japan}

\begin{abstract}
Making use of the semiclassical periodic orbit theory (POT), we
propose, for the first time, a method to exclusively evaluate the
shell effects associated with each of the nascent fragments
(prefragments) generated by the neck formation in nuclear fission
processes.  In spite of the strong indication of such shell effects in
asymmetric fragment mass distributions, they could not have been
accessed by any previous theoretical approach since most of the
single-particle wave functions are delocalized.  In the POT, we have
found that the prefragment shell effects can be naturally and
unambiguously identified as the contributions of the classical
periodic orbits localized in each of the prefragments.
For a numerical test, simple cavity potential models are employed with
the shape described by the three-quadratic-surface shape
parametrization.  Deformed shell energies are studied with the trace
formula for degenerate orbits in a truncated spherical cavity which
was recently derived [K.~Arita, Phys. Rev. C {\bf 98}, 064310 (2018)].
In this simple model, it is
shown that the prefragment shell effect dominates the total shell
energy shortly after the neck formation, and the magicity of the
heavier prefragment plays a significant role in establishing the
fission saddle with asymmetric shape which leads to an asymmetric
scission.
\end{abstract}
\pacs{%
03.65.Sq, 
21.60.-n, 
25.85.-w  
}
\maketitle

\section{Introduction}

In low-energy fissions of actinide nuclei, observed fragment
mass distributions show apparent asymmetry\cite{Andreyev2018}.  Since
symmetric shapes are always favored in the liquid-drop model
(LDM), the above asymmetry is attributed to the quantum shell effect.
For fission of nuclei in the range of proton number $Z=\mbox{90--100}$
and neutron number $N=\mbox{130--150}$, distributions of the heavier
fragments show distinctive concentrations at mass numbers around
$A\approx 140$, which is close to the doubly magic nucleus
${}^{132}\mbox{Sn}$.  From this observation, shell effects of the
fragments, which lower the binding energies of the final states, are
considered to play an important role as the origin of the asymmetric
fission.  In the fission of lighter nuclei, fragment distributions
are expected to be more symmetric because of the absence of such a
strong shell effect.  However, fissions of some neutron-deficient
mercury isotopes have been revealed to be asymmetric in recent
$\beta$-delayed fission experiments\cite{Andreyev2012,Prasad2015}.

Several theoretical investigations have been done to understand the
origin of these asymmetric fissions, e.g., with the shell correction
method in five-dimensional deformation
space\cite{Moller2012,Ichikawa2012} and with the fully microscopic
constrained Hartree-Fock-Bogoliubov (HFB)
method\cite{Warda2012,McDonnell2013}.  The interpretations of the
origins of the asymmetry are somewhat different among those
theoretical approaches, but the common recognition is that the shell
effect in the process of fission, which is responsible for the
positions of saddles in deformation space, plays an essential role in
determining the fragment mass distribution.

The fissioning nucleus gets more elongated in one direction, and a
neck is built which gradually separates the total system into two
subsystems.  Then, one may expect emergence of a kind of the shell
effect associated with those nascent fragments.  Following \
\cite{Schunck2016,Sadhukhan2017,Schmidt2018}, we call such a nascent
fragment a ``prefragment'' for brevity.

The effect of the prefragment shell structure was studied in the
two-center shell model\cite{Mosel_NPA,Mosel_PRC}.  The effect could be
seen immediately after getting over the second barrier of the
potential where the prefragments are not yet well separated.  In
modern microscopic HFB calculations\cite{Warda2012,McDonnell2013}, it
was found that the prefragment part of the density distribution is
quite similar to that of the independent spherical magic nucleus.
More recently, the spatial nucleon localization functions were
investigated to study the prefragment formation in nuclear
fission\cite{Zhang2016,Sadhukhan2017}.  These theoretical results
strongly indicate the existence of the prefragment shell effect, but
the physical mechanisms for their appearance have not been understood
well.

In this paper, we propose a simple way of extracting the prefragment
shell effect,
which originates from the spatially localized part of the
mean-field potential constricted in the middle.  Such shell effect
could not have been evaluated in any previous theoretical approach due
to the delocalization of the single-particle wave functions.  Focusing
on the fact that the semiclassical level density is expressed by a sum
of contributions of the classical periodic orbits, we found that the
prefragment shell effect can be naturally and unambiguously identified
as contributions of the orbit confined in each of the prefragments.
In Sec.~\ref{sec:theory}, after some remarks on the semiclassical
theory of shell structure, the idea of extracting prefragment shell
effect in the fission process is presented by means of the
periodic-orbit theory (POT).  In Section~\ref{sec:model}, properties
of the cavity potential model with three-quadratic-surfaces (TQS)
parametrization is discussed.  Semiclassical analysis of the TQS
cavity model with the POT is presented in Sec.~\ref{sec:application},
and Sec.~\ref{sec:summary} is devoted to a summary and discussions.

\section{Semiclassical foundation of the prefragment shell effect}
\label{sec:theory}

\subsection{Periodic orbit theory}
\label{sec:pot}

In the semiclassical POT, the single-particle energy level density
\begin{equation}
g(e)=\sum_i\delta(e-e_i)
\end{equation}
is represented as the sum over contributions of classical periodic
orbits\cite{Gutz72,BBText},
\begin{equation}
g(e)=\bar{g}(e)+\sum_\beta A_\beta(e)\sin\left[\frac{1}{\hbar}
 S_\beta(e)-\frac{\pi}{2}\mu_\beta\right]. \label{eq:trace_g_e}
\end{equation}
Here, $\bar{g}(e)$ represents the average level density which is
generally a smooth, monotonic function of $e$.  The second term
represents the oscillating part, and the sum is taken over all the
classical periodic orbits.  $S_\beta=\oint_\beta\bp\cdot d\br$ is the
action integral along the orbit $\beta$, $\mu_\beta$ is the Maslov
index related to the number of caustic points along $\beta$, and the
amplitude $A_\beta$ is determined by the period, degeneracy and
stability of the orbit.  This semiclassical expression
(\ref{eq:trace_g_e}) is known as the trace formula, and is derived
by evaluating the trace of the transition amplitude represented in the
path-integral form.

For the hard-wall potentials in which the action integral $S_\beta$ is
expressed as the product of the momentum $p=\hbar k$ and the orbit
length $L_\beta$, one has
\begin{gather}
g(k)=g(e)\frac{de}{dk}=\bar{g}(k)+2R_0\sum_\beta (kR_0)^{D_\beta/2}
 A_\beta \nonumber \\
 \times \sin\left(kL_\beta-\tfrac{\pi}{2}\mu_\beta\right).
\label{eq:trace_ld}
\end{gather}
$R_0$ is the length unit which is taken here as the radius of a
sphere having the same volume as the interior region of the cavity,
which is assumed to be conserved.  $D_\beta$ is the degeneracy
parameter for orbit $\beta$ which forms a $D_\beta$-parameter family
under the existence of a continuous symmetry.

Thus, the contribution of each orbit gives a regularly oscillating
function of $k$.  The scale of the oscillating structure is given by
the variation $\delta k$, which causes the change of phase by $2\pi$,
namely,
\begin{equation}
L_\beta \delta k\simeq 2\pi, \quad
\delta k \simeq \frac{2\pi}{L_\beta}.
\end{equation}
This relation tells us that the gross shell effect (with large $\delta
k$) is associated with the contributions of short orbits having
small $L_\beta$.  Amplitude is proportional to $k^{D_\beta/2}$ for a
degenerate $D_\beta$-parameter family of orbits, and the families with
higher degeneracies make more important contribution in the
semiclassical limit.  The dimensionless energy-independent amplitude
factor $A_\beta$ in Eq.~(\ref{eq:trace_ld}) is related to the
amplitude $A_\beta(e)$ in Eq.~(\ref{eq:trace_g_e}) by
\begin{equation}
A_\beta(e)\frac{de}{dk}=2R_0(kR_0)^{D_\beta/2}A_\beta.
\end{equation}
In the following, we use the symbol $A_\beta$ for the dimensionless
energy-independent amplitude, and the energy-dependent amplitude will
be referred to explicitly with the argument when necessary.

The analytic expression of the trace formula for the spherical cavity
was derived by Balian and Bloch\cite{BaBlo3}.  The families of regular
polygon orbits in the spherical cavity are classified by the two
integers $(p,t)$, where $p$ is the number of vertices and $t$ is the
number of turns around the center.  The length of the orbit $(p,t)$ is
given by
\begin{equation}
L_{pt}=R_0l_{pt},\quad l_{pt}=2p\sin\varphi_{pt},\quad
\varphi_{pt}=\frac{t\pi}{p}, \label{eq:len_sph}
\end{equation}
where $R_0$ is the radius of the cavity.  Polygon orbits $(p>2t)$ form
three-parameter families $(D_{p>2t}=3)$ generated by the
three-dimensional rotations, and diameter orbits $(p=2t)$ form
two-parameter families $(D_{p=2t}=2)$ since the rotation about the
orbit itself generates no family.  The contribution of the family
$(p,t)$ to the semiclassical level density is given by
\begin{gather}
g_{pt}^{\rm(sph)}(k)=2R_0(kR_0)^{D_{pt}/2}A_{pt}^{\rm(sph)}
\sin\left(kL_{pt}-\tfrac{\pi}{2}\mu_{pt}^{\rm(sph)}\right),
\end{gather}
with the amplitude factor
\begin{equation}
A_{pt}^{\rm(sph)}=\left\{\begin{array}{l@{\quad}l}
\sin 2\varphi_{pt}\sqrt{\dfrac{\sin\varphi_{pt}}{\pi p}} & (p>2t), \\
\dfrac{1}{2\pi t} & (p=2t),
                         \end{array}\right.
\end{equation}
and the Maslov index
\begin{equation}
\mu_{pt}^{\rm(sph)}=\left\{\begin{array}{l@{\quad}l}
2t-p-\frac32 & (p>2t), \\
2 & (p=2t).
                           \end{array}\right.
\end{equation}

Using the formula (\ref{eq:trace_g_e}) or (\ref{eq:trace_ld}), one can
derive the trace formula for the shell energy
\begin{align}
&\delta E(N)=\sum_\beta\frac{\hbar^2}{T_\beta^2}
A_\beta(e_F)\sin\left(\tfrac{1}{\hbar}S_\beta(e_F)
-\tfrac{\pi}{2}\mu_\beta\right) \nonumber \\
&\quad =\sum_\beta\frac{2\hbar^2(k_FR_0)^{1+D_\beta/2}}{ML_\beta^2}
A_\beta\sin\left(k_FL_\beta-\tfrac{\pi}{2}\mu_\beta\right),
\label{eq:trace_sce}
\end{align}
where $M$ is the particle mass and $e_F=(\hbar k_F)^2/2M$ is the Fermi
level corresponding to the particle number $N$.  Due to the extra
factor proportional to $T_\beta^{-2}\propto L_\beta^{-2}$, relative
contributions of the longer orbits are suppressed and only a few
shortest orbits dominate the shell energies in general.

Supposing that a certain orbit $\beta$ (isolated or degenerate)
dominates in the periodic-orbit sum (\ref{eq:trace_sce}), namely,
\begin{equation}
\delta E(N) \approx \frac{\hbar^2}{T_\beta^2} A_\beta(e_F)
 \sin\left(\tfrac{1}{\hbar}S_\beta(e_F)
 -\tfrac{\pi}{2}\mu_\beta\right),
\end{equation}
the shell energy would take the minima at
\begin{equation}
S_\beta(e_F)=2\pi\hbar\left(n+\tfrac{\mu_\beta-1}{4}\right)
\end{equation}
with integer $n$, and the corresponding particle number $N_n(e_F)$
gives the magic numbers for a given potential.  When considering the
shell structure as function of the deformation parameter $\delta$, the
formula
\begin{equation}
S_\beta(e_F,\delta)=2\pi\hbar\left(n+\tfrac{\mu_\beta-1}{4}\right)
\label{eq:cac}
\end{equation}
gives the condition for the parameter $\delta$ to lower the
deformation energy for particle number $N(e_F)$.

Strutinsky \etal\ have used the above constant-action lines
(\ref{eq:cac}) to explain the ridge-valley structure of the shell
energy in the $(\delta,N)$ plane, where $\delta$ is the quadrupole
deformation parameter\cite{Strut}.  Frisk discussed the origin of the
deformed shell structures related to the prolate-oblate asymmetry in
the spheroidal cavity model by means of constant-action-lines
analyses\cite{Frisk}.  It is also examined in the mean-field potential
model with more general radial dependence\cite{Arita2012}.

Brack \etal\ considered the origin of the asymmetric fission with the
POT\cite{Brack97}.  They analyzed the deformed shell effect in the
axially symmetric cavity potential model as function of the elongation
parameter $c$ and the asymmetry parameter $\alpha$, and found that the
fission path in the potential energy surface which connects the
minimum at the symmetric shape and the strongly elongated asymmetric
shape is simply explained by the constant-action line (\ref{eq:cac})
for the shortest periodic orbit in the $(c,\alpha)$ plane, which is
rewritten for the cavity model as
\begin{equation}
k_F L_\beta(c,\alpha)=2\pi\left(n+\tfrac{\mu_\beta-1}{4}\right).
\label{eq:cac_cavity}
\end{equation}

In the following, we consider the origin of asymmetric fission through
its relation to the prefragment shell effect.  For this aim, we
consider the way of extracting the contribution of spatially localized
prefragment to the shell effect by using the POT.

\subsection{Prefragment shell effect in the POT}
\label{sec:shell_frag}

Let us consider the deformation of a nucleus in a fission process where
it gets elongated.  In the LDM, one has competition
between the surface energy and the Coulomb energy.  With increasing
elongation, the Coulomb effect surpasses the surface effect and the
saddle is formed.  The shell effect, namely, the quantum fluctuation
is superimposed on this saddle structure, which typically constructs
a double-humped structure for nuclei in the actinide region.  Since
the Coulomb energy prefers a dumbbell shape rather than a rugby-ball
shape, the neck is formed after getting over the saddle.  One will expect
that the prefragments would acquire identities and begin to affect the
whole system after the formation of the neck.  It is also expected
that the above neck formation could be prompted by the quantum shell
effect.  However, in a purely quantum approach with the mean-field
potential, it is usually impossible to extract the shell effect
associated with the spatially localized part of the potential because
most of the single-particle wave functions are delocalized.

Here, we pay attention to the fact that the neck formation gives birth
to the classical periodic orbits confined in either of the prefragment
parts, which we call the prefragment orbits.  According to the spatial
localization characteristics of the orbits, we decompose the trace
formula (\ref{eq:trace_sce}) into three parts as
\begin{equation}
\delta E(N)=\delta E_1(N)+\delta E_2(N)+\delta E_3(N),
\label{eq:sh_frag}
\end{equation}
where $\delta E_1$ and $\delta E_2$ are contributions of the
prefragment orbits in the first and second prefragments, respectively,
and $\delta E_3(N)$ is the contribution of the other orbits traveling
between the two prefragments or staying in the neck part.  The
contribution of the prefragment periodic orbits can be naturally
identified as the shell effect associated with the prefragment part of
the potential.  This gives a simple and clear definition of the
prefragment shell effect.

\section{Cavity model with the three-quadratic-surfaces parametrization}
\label{sec:model}

\subsection{Shape parametrization for the fission process}

Since our aim is to understand the role of the prefragments in the
shell effect of fissioning nuclei, it is important to use a shape
parametrization which allows flexible control of the prefragment
deformations.  The three-quadratic-surfaces (TQS) parametrization is
one of the appropriate parametrizations for this purpose\cite{Nix69}.
In this parametrization, two horizontally aligned ellipsoids are
smoothly connected by the neck part which is also given by a quadratic
surface:
\begin{gather}
\rho^2(z)=\left\{\begin{array}{l@{\quad}l}
a_1^2-\dfrac{a_1^2}{c_1^2}(z-l_1)^2 & (l_1-a_1\leq z\leq z_1) \\[\baselineskip]
a_2^2-\dfrac{a_2^2}{c_2^2}(z-l_2)^2 & (z_2\leq z\leq l_2+a_2) \\[\baselineskip]
a_3^2-\dfrac{a_3^2}{c_3^2}(z-l_3)^2 & (z_1\leq z\leq z_2)
                 \end{array}\right.,
\label{eq:tqs}
\end{gather}
Among the eleven constants $\{a_{1-3},c_{1-3},l_{1-3},z_{1,2}\}$, six
are constrained by imposing the continuities of $\rho(z)$ and
$\rho'(z)$ at the joints $z=z_1, z_2$, the center-of-mass condition,
and the volume-conservation condition.  Thus, one eventually has five free
parameters to determine the shape of the potential.

We use the set of parameters $\{\sigma_{1-3},\alpha_{1-3}\}$ defined
in \cite{Nix69} as
\begin{align}
\sigma_1&=\frac{l_2-l_1}{u}, \quad
\alpha_1=\frac{l_1+l_2}{2u} \quad
\left(u=\sqrt{\frac{a_1^2+a_2^2}{2}}\right), \nonumber \\
\sigma_2&=\frac{a_3^2}{c_3^2}, \quad
\alpha_2=\frac{a_1^2-a_2^2}{u^2} \nonumber \\
\sigma_3&=\frac12\left(\frac{a_1^2}{c_1^2}+\frac{a_2^2}{c_2^2}\right), \quad
\alpha_3=\frac{a_1^2}{c_1^2}-\frac{a_2^2}{c_2^2}.
\end{align}
$\sigma_1$ is proportional to the distance between the centers of the
two prefragments and is regarded as the elongation parameter.
$\sigma_2$ is proportional to the curvature of the middle surface and
controls the shape of the neck.  For the fission deformations, the neck
surface is usually concave and $\sigma_2$ takes negative values.
$\alpha_2$ is related to the prefragment mass asymmetry.  $\sigma_3$
and $\alpha_3$ control the deformation of the two prefragments.  The
above five are taken as the free shape parameters.  The parameter
$\alpha_1$ which describes the asymmetry of the centers of the
prefragments is automatically determined by the above five parameters.

In this work, we fix the shapes of both prefragments to be
spherical ($\sigma_3=1,\alpha_3=0$), and also fix the value of the
neck parameter to a typical value $\sigma_2=-0.6$, found in
realistic calculations of the fission paths for U and Pu
isotopes\cite{Ichikawa2012}.  The radius of the $j$th prefragment
$(j=1,2)$ is denoted by $R_j\;(=a_j=c_j)$ below.
\begin{figure}
\includegraphics[width=\linewidth]{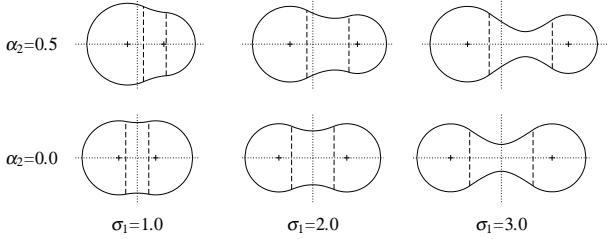}
\caption{\label{fig:shape_TQS}
The shapes of the potential surfaces in the TQS parametrization
varying the elongation parameter $\sigma_1$ and the prefragment mass
asymmetry parameter $\alpha_2$, with spherical prefragments
($\sigma_3=1$, $\alpha_3=0$) and fixed value of the neck parameter
$\sigma_2=-0.6$.  The broken lines represent the joints of the
neighboring quadratic surfaces, and the cross symbols represent the
centers of the left and right prefragments.  Horizontal and vertical
dotted lines indicate the symmetry axis and the position of the center
of mass, respectively.}
\end{figure}
The shapes of the potential surface for several values of
$(\sigma_1,\alpha_2)$ are displayed in Fig.~\ref{fig:shape_TQS}.

\begin{figure}
\includegraphics[width=\linewidth]{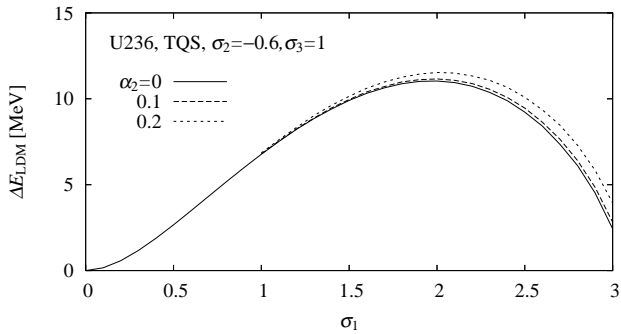}
\caption{\label{fig:ldm}
Liquid-drop model deformation energies
$\varDelta E_{\rm LDM}$ for $^{236}\mbox{U}$ nucleus
plotted as functions of the elongation parameter $\sigma_1$ with
fixed values of the mass asymmetry parameter $\alpha_2$.
Solid, long-dashed and short-dashed
lines represent the results for $\alpha_2=0$, $0.1$ and $0.2$,
respectively.}
\end{figure}

Figure~\ref{fig:ldm} shows the LDM deformation energies for the
$^{236}\mbox{U}$ nucleus,
\begin{equation}
\varDelta E_{\rm LDM}
=(B_S-1)a_S A^{2/3}+(B_C-1)a_C \frac{Z^2}{A^{1/3}},
\end{equation}
as functions of the elongation parameter $\sigma_1$ with some fixed
values of the mass asymmetry parameter $\alpha_2$.  $a_S$ ($a_C$) is
the LDM surface (Coulomb) parameter for spherical nuclei, and $B_S$
($B_C$) is the deformation factor\cite{RSText} (see the
Appendix~\ref{sec:app} for $B_S$).  One finds the saddle at
$\sigma_1\simeq 2.0$ for this nucleus, and also for nuclei in the
actinide regions. For any given value of $\sigma_1$, the LDM energy
takes the minimum at the symmetric shape, $\alpha_2=0$.  Therefore, a
trend to asymmetric shapes is considered to be purely a quantum effect in
this TQS parametrization.

\subsection{Single-particle spectra}

The mean-field potential for a heavy nucleus is approximately flat
near the center with depth of about $50$~MeV, and sharply approaches
zero around the surface with diffuseness of about $0.7$~fm.  The
existence of strong spin-orbit coupling and pairing correlations is
well known, but here we focus our attention on the roles of the
shape evolution and adopt a simple infinite-well (cavity) potential
model.  The single-particle eigenvalue problem is solved by the
spherical-wave decomposition method (SWDM)\cite{MP95}.
For the wave function expressed as the superposition of the spherical
waves with given energy and $K$ quantum number, one obtains the energy
eigenvalue at which the boundary condition is satisfied extremely well.

\begin{figure}[t]
\centering
\includegraphics[width=.9\linewidth]{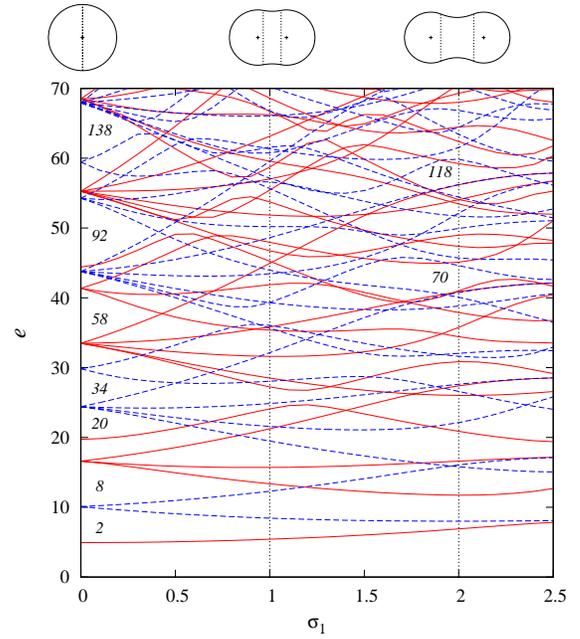}
\caption{\label{fig:nils_s} 
Single-particle level diagram for the symmetric TQS cavity model as
function of the elongation parameter $\sigma_1$.  Solid (red) and
broken (blue) lines represent the levels with positive and negative
parities, respectively.  The shapes of the potential surface at
$\sigma_1=0$, $1.0$, and $2.0$ are shown at the top of the diagram.}
\end{figure}

Figure~\ref{fig:nils_s} shows the level diagram for a symmetric
deformation.  $\sigma_1=0$ corresponds to the spherical shape, and one
sees shell closures for magic numbers $N=2, 8, 20, 34, 58, 92, 138,
\cdots$.  Deformed shell closures are also found at large
deformations, e. g. $N=70, 118$ at $\sigma_1\simeq 2$.  One will also
see some of the levels with different parities approaching each
other at large $\sigma_1$, which may be related to localization of
single-particle wave functions in each of the prefragments.

\begin{figure}
\centering
\includegraphics[width=.9\linewidth]{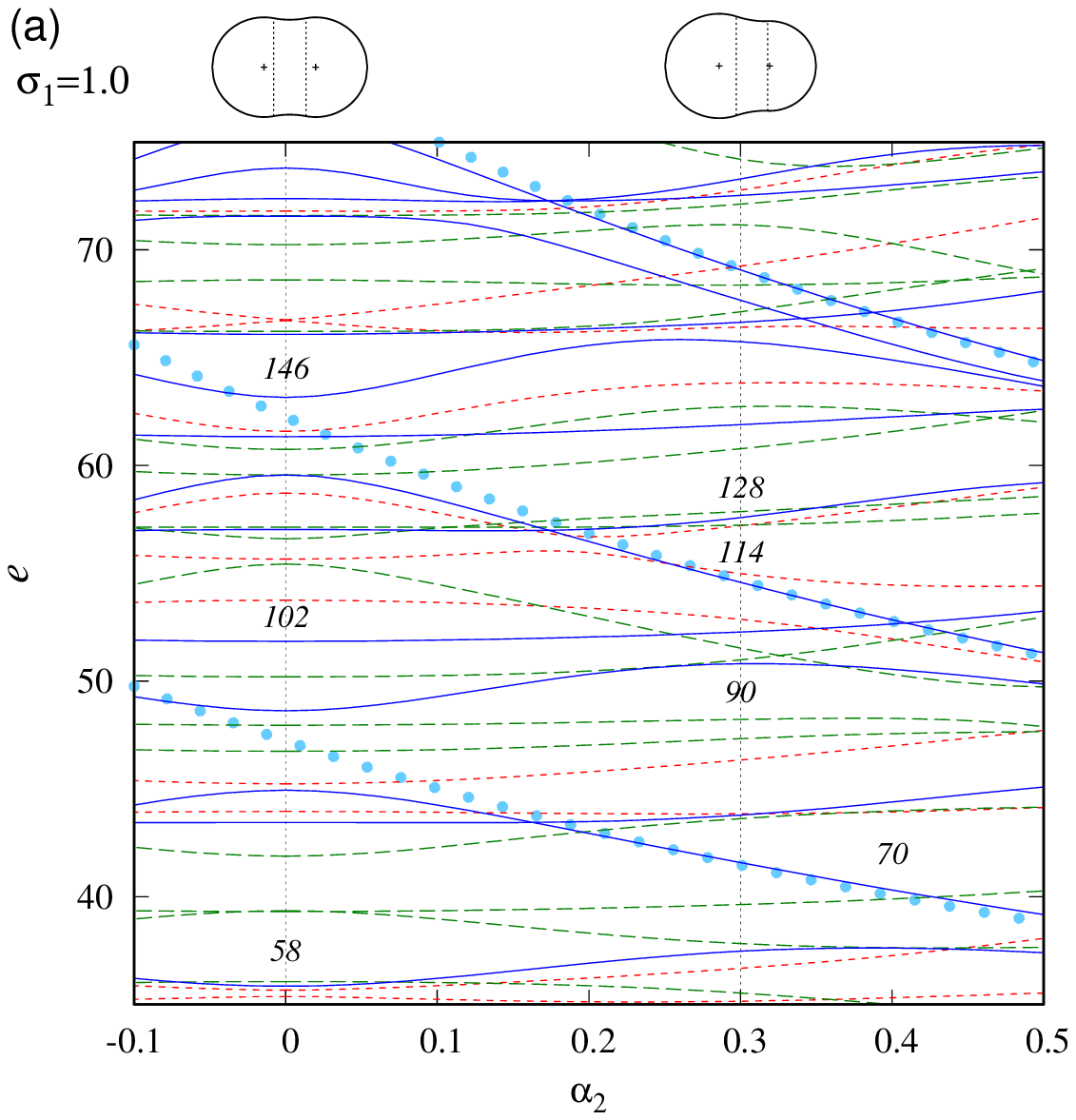} \\
\includegraphics[width=.9\linewidth]{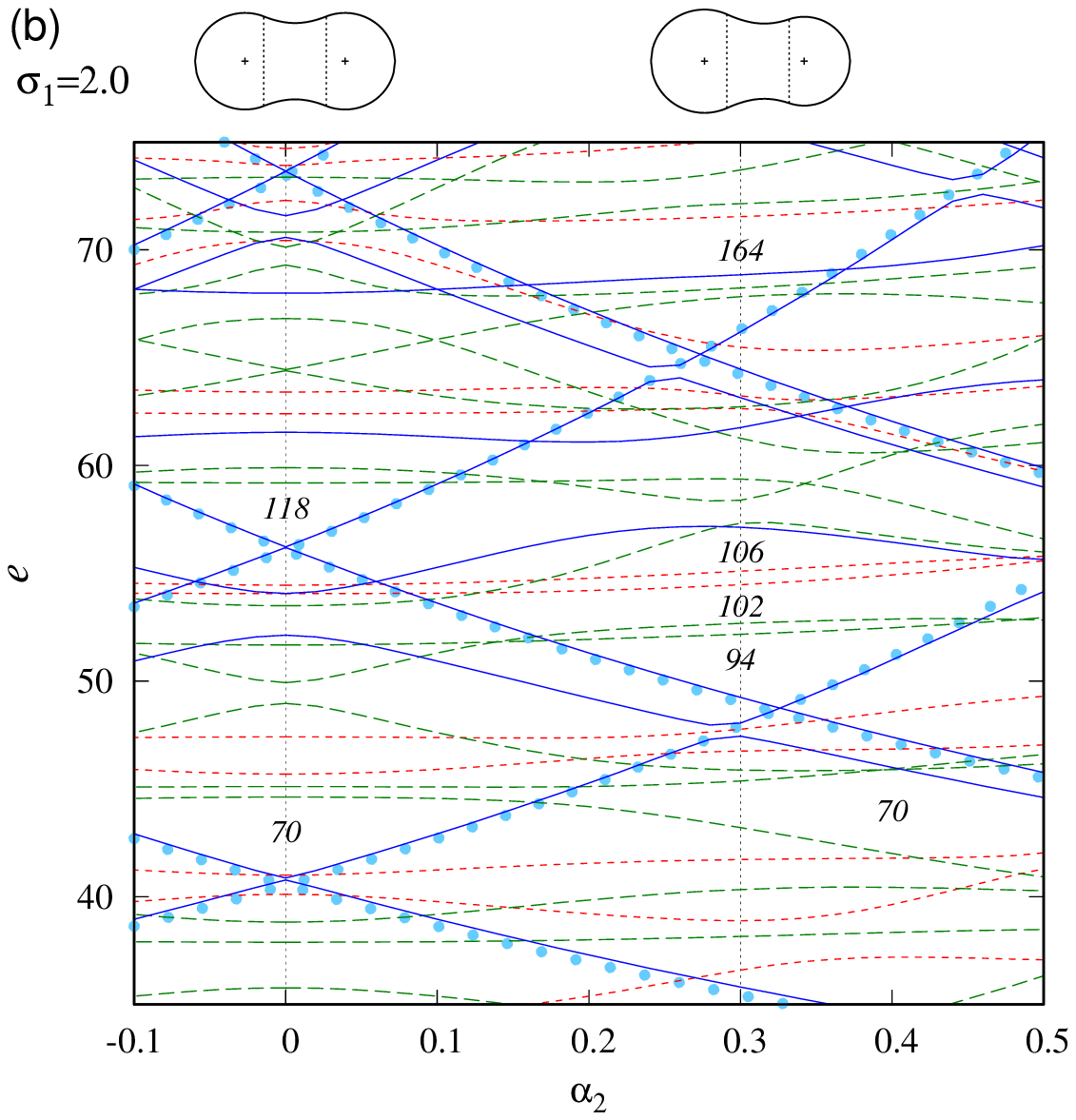}
\caption{\label{fig:nils_a} 
Single-particle level diagrams as functions of the asymmetry parameter
$\alpha_2$, with the elongation parameter $\sigma_1=1.0$ (a) and 2.0
(b).  The short-dashed (red), long-dashed (green) and solid (blue) lines
represent levels with the magnetic quantum numbers $K=0$, $1\leq K\leq
2$, and $K\geq 3$, respectively.  The thick dotted (light blue) lines are
drawn so as to be proportional to $1/R_j^2$, which would be helpful for
recognizing the levels indicating localization to each of the
prefragments.  Shapes of the surface at $\alpha_2=0$ and $0.3$ are
displayed at the top of each panel.}
\end{figure}

In Fig.~\ref{fig:nils_a}, level diagrams as functions of the asymmetry
parameter $\alpha_2$ with fixed values of the elongation parameter
$\sigma_1$ are shown.  One finds significant differences between the
diagrams for $\sigma_1=1.0$ and $2.0$.  For $\sigma_1=1.0$, most of
the levels are nearly horizontal as functions of $\alpha_2$ for small
$\alpha_2$, which indicates a chaotic nature.  For large $\alpha_2$,
one sees some high-$K$ levels going down with increasing $\alpha_2$.
Since their slopes are almost proportional to $1/R_1^2$ indicated by
the thick dotted lines, they are considered to be the states localized in
the heavier prefragments whose radius $R_1$ increases with $\alpha_2$.
In the diagram for $\sigma_1=2.0$ one sees both descending and
ascending high-$K$ levels with avoided crossings whose slopes are
almost proportional to $1/R_1^2$ and $1/R_2^2$, respectively.  Those
levels indicate the appearance of the states localized in each of the
prefragments due to the developed neck\cite{Brack2001}.  One finds
several shell gaps showing up near the crossing points around
$\alpha_2\approx 0.3$.

The SWDM works sufficiently well for small elongation, but suddenly
turns inaccurate for $\sigma_1>2.5$.  We will have to find another
method to look at the behavior of the level structure up to the
scission point, which remains as a future problem.

\subsection{Classical periodic orbits}

\begin{figure}
\includegraphics[width=\linewidth]{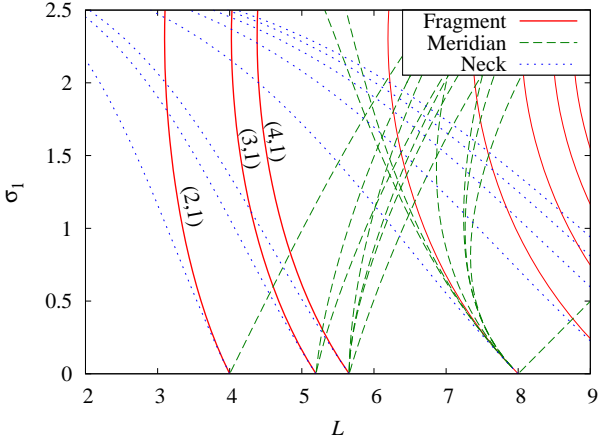}
\caption{\label{fig:po_len} 
Lengths of the symmetric classical periodic orbits in the TQS cavity
model as functions of the elongation parameter $\sigma_1$, with
spherical prefragments and the neck parameter fixed at
$\sigma_2=-0.6$.  Solid lines (red) represent the prefragment orbits
$n(p,1)$, broken lines (green) represent the meridian-plane orbits,
and dotted lines (blue) represent the equatorial regular polygon
orbits in the neck surface (see Ref.~\cite{Part1}).}
\end{figure}

In the TQS cavity model under consideration, classical periodic orbits
are classified into the following four categories:
\begin{enumerate}\def\labelenumi{(\roman{enumi})}\itemsep=0pt
\item prefragment polygon orbits in either of the spherical
prefragments, which form three-parameter families;
\item prefragment diameter orbits in either of the spherical
prefragments, which form two-parameter families;
\item meridian-plane orbits, equatorial orbits in the neck surface
and three-dimensional orbits, which form one-parameter
families generated by the rotation about the symmetry axis
\item an isolated straight-line orbit along the symmetry axis;
\end{enumerate}
Figure.~\ref{fig:po_len} shows the lengths of some classical periodic
orbits in a symmetric TQS cavity ($\alpha_2=0$) plotted as functions
of the elongation parameter $\sigma_1$.
Since the orbit with higher degeneracies play a dominant role for the
shell effect, we might be able to consider the shell energy only with
the contributions of the orbit families belonging to categories
(i) and (ii), at least for sufficiently large $\sigma_1$.  The
prefragment orbit families in each of the spherical prefragments are
labeled by the two indices $(p,t)$ just as in an isolated spherical
cavity discussed in Sec.~\ref{sec:pot}.  Polygon orbits $p>2t$ belong
to (i) and diameter orbits $p=2t$ to (ii).  The length of the orbit
$(p,t)$ in the $j$th prefragment $(j=1,2)$ is given by
\begin{equation}
L_{j,pt}=R_jl_{pt},
\end{equation}
with $l_{pt}$ defined by Eq.~(\ref{eq:len_sph}).  For small elongation
$\sigma_1$, prefragment orbits exist only in the narrow region around
the planes $z=l_1$ and $z=l_2$, but the region broadens as the neck
develops with increasing $\sigma_1$, and then their contributions to
the level density will become more important.  As we discussed in
\cite{Part1}, the orbit families (i) and (ii) are accompanied by the
marginal orbits having lower degeneracies than the principal family.
Their contributions to the semiclassical level density should be
treated separately from those of the principal families, and they play
significant roles especially at small elongation.

\section{Semiclassical analysis of the prefragment shell effect}
\label{sec:application}

\subsection{Fourier transforms of the level density}

Thanks to the simple momentum dependence of the action integral, one
can obtain clear correspondence between quantum shell structure and
classical orbits through the Fourier transformation of the level
density.  Let us define the Fourier transform of the level density
$g(k)$ with respect to the wave number $k$ by
\begin{equation}
F(L)=\sqrt{\frac{2}{\pi}}\int_0^\infty dk\,
 g(k)e^{ikL}e^{-\frac12(k/k_c)^2}. \label{eq:fourier}
\end{equation}
The Gaussian in the integrand is introduced to truncate the high
energy part $(k\gg k_c)$ of the spectrum which is unavailable in the
numerical calculations.  The cutoff momentum $k_c$ is taken as
$k_c\approx k_{\rm max}/\sqrt{2}$, $k_{\rm max}$ being the value where
the average density of numerical quantum levels begins to deviate from
the (extended) Thomas-Fermi value.  Inserting the quantum level
density $g(k)=\sum_i\delta(k-k_i)$, one has
\begin{equation}
F^{\rm qm}(L)=\sqrt{\frac{2}{\pi}}\sum_{i=1}^\infty e^{ik_i L}
e^{-\frac12(k_i/k_c)^2},
\label{eq:ftl_qm}
\end{equation}
which can be easily calculated from the quantum spectrum.  Inserting
the semiclassical level density (\ref{eq:trace_ld}) into
(\ref{eq:fourier}), one obtains
\begin{align}
F^{\rm sc}(L)=F_0(L) &+\sum_\beta(k_cR_0)^{1+D_\beta/2}
A_\beta e^{-i\pi\mu_\beta/2} \nonumber \\
&\times \Lambda_{D_\beta}\left(\tfrac{L-L_\beta}{\gamma}\right),
\quad \gamma=k_c^{-1}, \label{eq:ft_cl}
\end{align}
where $\Lambda_D(y)$ defined by
\begin{equation}
\Lambda_D(y)=\sqrt{\frac{2}{\pi}}\int_0^\infty dx x^{D/2}e^{ixy}e^{-x^2/2}
\end{equation}
is the function having a single peak at $y=0$ with height
$|\Lambda_D(0)|\simeq 0.8$ and width $\varDelta y\simeq 4$ for $D\geq
1$\cite{Part1}.  Accordingly, (\ref{eq:ft_cl}) makes a function
exhibiting peaks at the lengths of the orbits, $L=L_\beta$, with width
$\varDelta L\simeq 4\gamma$ and height $|F(L_\beta)|\propto A_\beta$.
Thus, we can extract information on the periodic-orbit contributions
out of the quantum spectrum by means of the above Fourier analysis.

\begin{figure}
\includegraphics[width=\linewidth]{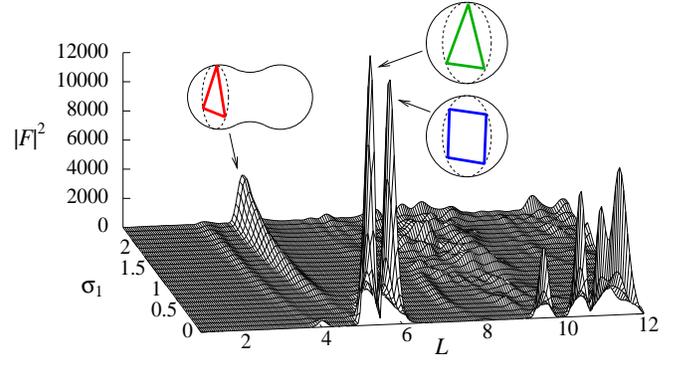}
\caption{\label{fig:fbird}
Fourier transform of the single-particle level density.  The squared
amplitude $|F^{\rm qm}(L;\sigma_1)|^2$ is plotted as a function of
$L$ and the elongation parameter $\sigma_1$.}
\end{figure}

Figure~\ref{fig:fbird} shows the moduli of the Fourier transform
(\ref{eq:ftl_qm}) as function of $\sigma_1$ and $L$ for symmetric
deformation ($\alpha_2=0$).  For the spherical shape ($\sigma_1=0$),
one sees prominent peaks at the equilateral triangle orbit
$L_{31}=3\sqrt{3} \simeq 5.19$ (in units of $R_0$, which also applies
below) and the square orbit $L_{41}=4\sqrt{2} \simeq 5.65$.  One also
sees a small peak at the diameter orbit $L_{21}=4$.  The polygon
orbits with more than five vertices have length close to that of
the square orbit and their peaks are not resolved in this calculation.
With increasing $\sigma_1$, the above two peaks promptly decay due to
the breakdown of the spherical symmetry, but one can see the peak
corresponding to the prefragment triangle family (3,1) growing for
larger $\sigma_1$ as discussed in the previous section.

\subsection{Shell energy}

Taking account of the effect of marginal orbits, the contribution of
the prefragment polygon orbit $(p,t)$ in the $j$th prefragment
$(j=1,2)$ is expressed as
\begin{align}
g_{j,pt}(k)&=2R_j\sum_{D=1}^3
(kR_j)^{D/2}A_{j,pt}^{(D)}\sin\left(kL_{j,pt}-\tfrac{\pi}{2}
\mu_{pt}^{(D)}\right) \nonumber \\
&=2R_j\Im\left[\mathcal{A}_{j,pt}(k)e^{ikL_{j,pt}}\right],
\label{eq:trace_jpt}
\end{align}
with the complex amplitude
\begin{equation}
\mathcal{A}_{j,pt}(k)=\sum_D(kR_j)^{D/2}A_{j,pt}^{(D)}e^{-i\pi
\mu_{pt}^{(D)}/2}.
\end{equation}
The summation is taken over the degeneracy parameter $1\leq D\leq 3$.
For the polygon orbit $(p>2t)$, the term for $D=3$ is the contribution
of the principal three-parameter family which is given by a certain
fraction of the formula for the spherical cavity.  $D=2$ is the
contribution of the marginal orbits which have one vertex on the joint
of the prefragment and neck surfaces.  $D=1$ is the contribution of
the secondary marginal orbits which have two vertices on the joint.
The last term is generally much smaller than the preceding two terms.
For the diameter family $(p=2t)$, there is no $D=3$ term and $D=2$ is
the contribution of the principal two-parameter family given by a
certain fraction of that of the spherical cavity.  $D=1$ is the
contribution of the marginal orbits.  The analytic expressions for the
amplitudes $A_{j,pt}^{(D)}$ and the Maslov indices $\mu_{pt}^{(D)}$
are given in Ref.~\cite{Part1}.

\begin{figure}
\centering
\includegraphics[width=\linewidth]{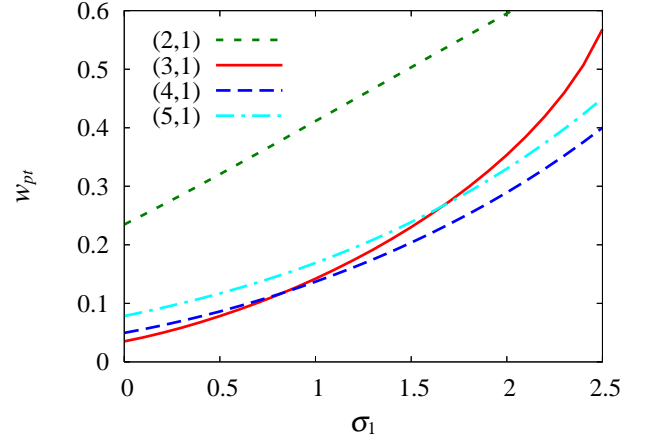}
\caption{\label{fig:zp}
The relative amplitude factors $\w_{j,pt}$, given by
Eq.~(\ref{eq:zp}), of some short prefragment orbit families $(p,t)$
for symmetric shapes as functions of the elongation parameter
$\sigma_1$.  The wave number is put to the Fermi level $k=k_F$
corresponding to $N=100$.}
\end{figure}

Defining the amplitude relative to that for the entire spherical cavity
\begin{equation}
\w_{j,pt}(k)=\frac{|\mathcal{A}_{j,pt}(k)|}{(kR_j)^{D_{pt}/2}A_{pt}^{\rm(sph)}}
\label{eq:zp}
\end{equation}
and the effective Maslov index $\mu^{\rm(eff)}$ given through
\begin{equation}
-\tfrac{\pi}{2}\mu_{j,pt}^{\rm(eff)}(k)=\arg\mathcal{A}_{j,pt}(k),
\end{equation}
one can rewrite Eq.~(\ref{eq:trace_jpt}) as
\begin{align}
g_{j,pt}(k)=&2R_j(kR_j)^{D_{pt}/2}\w_{j,pt}(k)A_{pt}^{\rm(sph)} \nonumber \\
&\times\sin\left(kL_{j,pt}-\tfrac{\pi}{2}\mu_{j,pt}^{\rm(eff)}\right),
\end{align}
and the contribution to the shell energy is expressed as
\begin{align}
\delta E_{j,pt}(N)=& \frac{2\hbar^2(k_FR_j)^{1+D_{pt}/2}}{ML_{j,pt}^2}
\w_{j,pt}A_{pt}^{\rm(sph)} \nonumber \\
&\times\sin\left(k_FL_{j,pt}-\tfrac{\pi}{2}\mu_{j,pt}^{\rm(eff)}
\right).  \label{eq:trace_tqs}
\end{align}
Note that $\w_{j,pt}$ and $\mu_{j,pt}^{\rm(eff)}$ depend on the
single-particle energy and are evaluated at the Fermi level.
Figure~\ref{fig:zp} shows the relative amplitudes $\w_{j,pt}$ at the
Fermi level corresponding to $N=100$ for several short-periodic-orbit
families.  At $\sigma_2\simeq 2.0$ corresponding to the saddle of the
LDM energy for actinide nuclei where the neck is expected to
grow considerably, the value of $\w_{31}$ for the most important
triangle orbit family amounts to 0.3--0.4, which means that the
prefragment effect is more than 30\% of the spherical shell effect
which is quite strong.

\begin{figure}
\includegraphics[width=\linewidth]{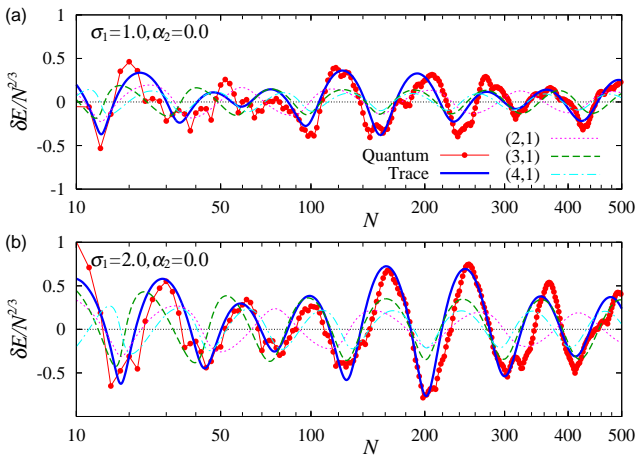}
\caption{\label{fig:trace_s} 
Shell energies $\delta E(N)$ as functions of particle number $N$ for
symmetric shapes $\sigma_1=1.0$ (a) and $2.0$ (b) with $\alpha_2=0$.
The abscissa is taken to be linear in $N^{1/3}$, which also applies
in the following
Figs.~\ref{fig:trace_a}--\ref{fig:sce_a}.  The thin solid lines with
dots (red) represent the quantum results, and thick solid lines (blue)
represent the results of the semiclassical trace formula with the
contributions of the prefragment orbit families.  The individual
contributions of the diameter (2,1), triangle (3,1), and square (4,1)
orbits are shown by the dotted (magenta), dashed (green), and
dash-dotted (cyan) lines, respectively.}
\end{figure}

The value of $\w_{pt}$ for the diameter is considerably larger than
those for polygons at small $\sigma_1$, which suggests that the
relative contribution of the diameter family is more important in the
truncated cavity than in the full spherical cavity.  However, its
absolute contribution is much less important than those of
three-parametric polygon families due to the smaller degeneracy.
Ignoring the contributions of orbits with smaller degeneracies
$D_\beta<2$, one can write the shell energy (\ref{eq:sh_frag}) only
with the prefragment orbit family contributions,
\begin{equation}
\delta E_j(N)=\sum_{pt}\delta E_{j,pt}(N) \quad (j=1,2).
\label{eq:trace_frag}
\end{equation}
Figure~\ref{fig:trace_s} shows the semiclassical trace formula $\delta
E(N)=\delta E_1(N)+\delta E_2(N)$ with Eqs.~(\ref{eq:trace_frag}) and
(\ref{eq:trace_tqs}) for symmetric deformations, $\sigma_1=1.0$ and
$2.0$ with $\alpha_2=0$.  Contributions of the prefragment orbit
families $(p,1)$ $(2\leq p\leq 5)$ are taken into account.  The orbit
in either the first or the second prefragment gives the equivalent
contribution for the symmetric shapes.  One sees that the quantum
results are nicely reproduced by our trace formula.  In the figure,
each contribution of (2,1), (3,1), and (4,1) orbits is also shown.
One can see that the main oscillating structure is governed by (3,1),
and the modulation of the shell structure, which is called a
supershell structure, is provided due to the interference between
different orbits.  For instance, one sees deep energy minima for
$N=200$ where the three orbits (3,1), (4,1), and (2,1) make
constructive contributions, while the minimum is shallower for $N=80$
where the above contributions are somewhat destructive.

\begin{figure}
\includegraphics[width=\linewidth]{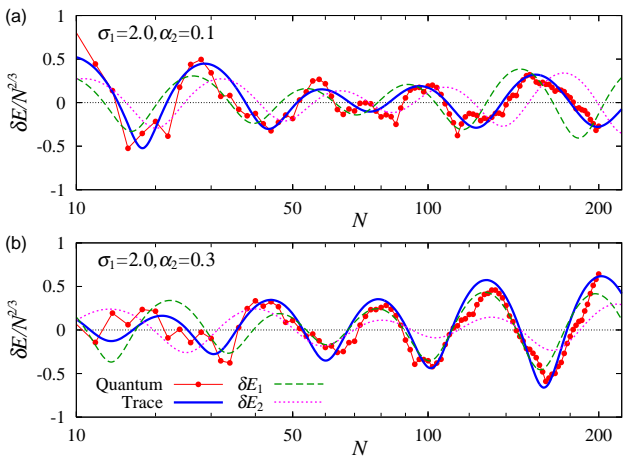}
\caption{\label{fig:trace_a} 
Shell energies $\delta E(N)$ as functions of particle number $N$ for
asymmetric shapes $\alpha_2=0.1$ (a) and $0.3$ (b) with fixed
elongation parameter $\sigma_1=2.0$.  The thin solid lines with dots (red)
represent quantum results and the thick solid lines (blue) represent the
results of semiclassical trace formula with contributions of the
prefragment orbit families.  The contributions of prefragment orbits
in the first (heavy) and second (light) prefragments are shown by the
dashed (green) lines and dotted (magenta) lines, respectively.}
\end{figure}

The results for asymmetric shapes, $\alpha_2>0$ with $\sigma_1=2.0$,
are shown in Fig.~\ref{fig:trace_a}, where the trace formula also
succeeds in reproducing the quantum results.  The contribution of each
prefragment is also shown in this figure.  Since the sizes of the two
prefragments are different for asymmetric shapes, the same types of orbits
but in different prefragments interfere with each other and bring
about another supershell effect.
For instance, in the panel for $\alpha_2=0.3$,
the shell effect around $N\approx 160$ is much more enhanced than the
other particle-number regions due to the constructive
contributions of the two prefragments.

\subsection{Shape asymmetry and the prefragment magicity}

As we have shown above, the shell energy is dominated by the
contributions of the prefragment orbit families.  The main oscillating
structure is governed by the triangle family (3,1).  Looking at
Fig.~\ref{fig:zp}, one will also note that the relative amplitudes
$\w_{pt}$ for the other polygon families take similar values through
the change of $\sigma_1$.  The Maslov indices for the marginal orbits
are found to be given by $\mu_{pt}^{(3)}=\mu_{pt}^{(2)}+\frac12
=\mu_{pt}^{(1)}+1$, and it may be allowed to approximate as
$\mu_{pt}^{\rm(eff)}\approx \mu_{pt}^{\rm(sph)}$.  Therefore, the
prefragment shell effect should be similar to that of the independent
spherical cavity.  These suggest the possibility of expressing the
shell energy of the total system containing the two spherical
prefragments in terms of those for two spherical cavities.

By replacing $\w_{j,pt}$ with $\w_{j,31}$ and $\mu_{j,pt}^{\rm(eff)}$
with $\mu_{j,pt}^{\rm(sph)}$ in Eq.~(\ref{eq:trace_tqs}), one obtains
the relation
\begin{align}
\delta E_j^{\rm(frag)}(N)&\approx \w_{j,31}\sum_{pt}A_{pt}^{\rm(sph)}(k_F)
\sin\left(k_FL_{j,pt}-\tfrac{\pi}{2}\mu_{j,pt}^{\rm(sph)}\right)
\nonumber \\
&=\w_{j,31}\delta E^{\rm(sph)}(N_j;R_j). \label{eq:sce_frag_sph}
\end{align}
$N_j$ is the particle number for the spherical cavity with the radius
$R_j$ corresponding to the Fermi level $k_F$, which will be referred to
as the particle number of the $j$th prefragment for convenience.
Since the particle number of the spherical cavity of radius $R$ is
proportional to $(k_FR)^3$ in the lowest order (see
Appendix~\ref{sec:app}), one has
\begin{gather}
N_j\approx \left(\frac{R_j}{R_0}\right)^3 N, \label{eq:n_frag}
\end{gather}
where $R_0$ is the radius of the total system in the spherical limit.
In the above approximation (\ref{eq:sce_frag_sph}), the diameter
contribution causes the main error due to the difference between
$\w_{21}$ and $\w_{31}$, which may not be important for large $N$ and
will decrease for large $\sigma_1$.  Consequently, the total shell
energy can be approximately represented in terms of the shell energies
of the two spherical cavities as
\begin{equation}
\delta E(N)\approx \sum_j \w_{j,31}\delta E^{\rm(sph)}(N_j;R_j),
\label{eq:esh_frag_sph}
\end{equation}
which might be smoothly linked to the scission point where the shell
energy is given by the sum of those for the two independent fragments.
Assuming the existence of the dominant prefragment orbit, the above
kind of relation is expected to hold in more generic mean-field
models, and it tells us how the prefragment shell effect grows in a
wide region of deformation from the second saddle to scission through
the factor $w$, representing the contribution of the dominant
prefragment orbit.

\begin{figure}
\includegraphics[width=\linewidth]{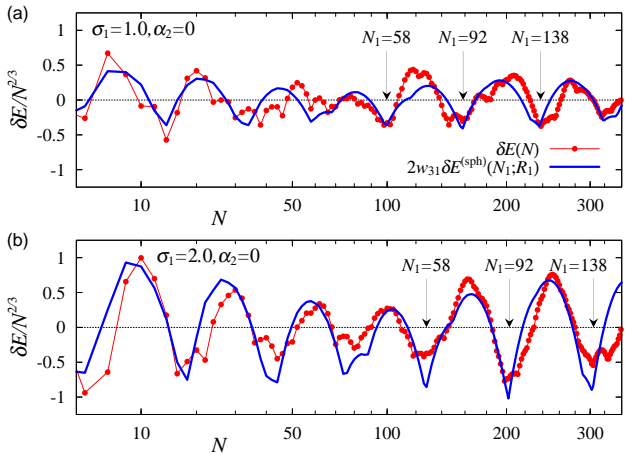}
\caption{\label{fig:sce_s} 
Shell energies as functions of the particle number $N$ for symmetric
deformations $(\alpha_2=0)$ with the elongation parameter
$\sigma_1=1.0$ (a) and 2.0 (b).  The thin line with dots (red) represents
the quantum result and thick line (blue) represents the prefragment
shell effect $\delta E_{\rm frag}(N)$ which is approximately evaluated
in terms of the shell energy of the spherical cavity as in
Eq.~(\ref{eq:esh_frag_sph}).  Arrows indicate the magic numbers of the
spherical prefragments.}
\end{figure}

\begin{table}
\caption{\label{tab:n_frag_s}
The prefragment radius $R_1(=R_2)$ in units of $R_0$, and the
prefragment particle number $N_1(=N_2)$ relative to the total particle
number $N$, used in evaluating the prefragment shell effect given by
Eq.~(\ref{eq:esh_frag_sph}) for symmetric deformations ($\alpha_2=0$)
at several values of the elongation parameter $\sigma_1$.}
\centering
\tabcolsep=8pt
\begin{tabular}{c|c|c}
\hline\hline
$\sigma_1$ & $R_1/R_0$ & $N_1/N$ \\ \hline
0.5 & 0.90033 & 0.72980 \\
1.0 & 0.83594 & 0.58414 \\
1.5 & 0.79417 & 0.50089 \\
2.0 & 0.76943 & 0.45552 \\
2.5 & 0.75979 & 0.43861 \\
3.0 & 0.76637 & 0.45011 \\
\hline\hline
\end{tabular}
\end{table}

In Fig.~\ref{fig:sce_s}, we show the shell energies evaluated using
Eq.~(\ref{eq:esh_frag_sph}) for symmetric shapes and compare them with
the exact ones.  We found that the major pattern of the shell energy
is nicely reproduced by the sum of the two shell energies of spherical
cavities for $\sigma_1\gtrsim 1.0$ with the coefficient $\w_{31}$
obtained by our formula.  Most of the shell energy minima are
explained by the prefragment magic numbers $N_j=58,92,138$ as
indicated in the figures.  The values of the prefragment radius $R_1$
relative to the radius $R_0$ of the total system in the spherical
limit and the corresponding prefragment particle number $N_1$ relative
to the total particle number $N$ are given in Table~\ref{tab:n_frag_s}
for symmetric shapes with several values of $\sigma_1$.  For small
$\sigma_1$, two provisional prefragment spheres overlap and $N_1/N$ is
greater than $1/2$.  With increasing $\sigma_1$, $N_1/N$ becomes
smaller than $1/2$, which suggests that the magicity of the
prefragments contribute to the shell-energy minima at $N>2N_1$.  For
instance, the $N\approx 200$ system has a deep energy minimum for
symmetric deformation with $\sigma_1=2.0$ which is related to the
prefragment magic number $N_1=92$.  This kind of relation might be
useful in considering the effect of the prefragment magic numbers on
the fragment mass distribution.

\begin{figure}
\includegraphics[width=\linewidth]{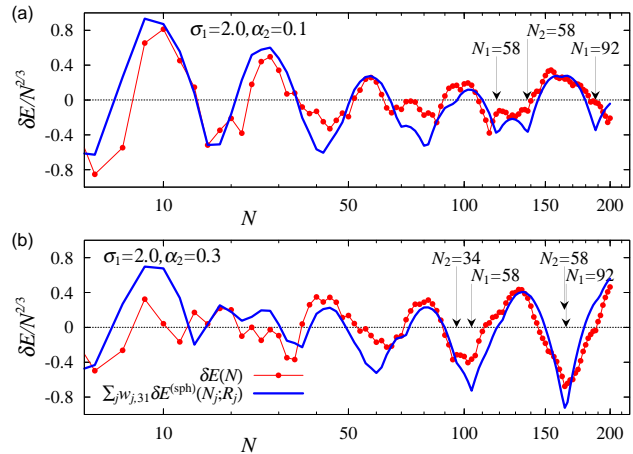}
\caption{\label{fig:sce_a} 
Same as Fig.~\ref{fig:sce_s} but for asymmetric shapes.
The mass asymmetry parameter is taken as $\alpha_2=0.1$ (a) and
$0.3$ (b), with the common elongation parameter $\sigma_1=2.0$.}
\end{figure}

In Fig.~\ref{fig:sce_a}, we examine the relation
(\ref{eq:sce_frag_sph}) for asymmetric shapes with the elongation
parameter fixed at $\sigma_1=2.0$.  The radii $R_j$ in units of $R_0$
and the prefragment particle numbers $N_j$ relative to the total
particle number $N$ for several values of $\alpha_2$ are listed in
Table~\ref{tab:n_frag_a}.  The exact shell energies are reproduced
quite well by the approximation (\ref{eq:sce_frag_sph}).  One obtains
especially large shell energy gains when the particle numbers of both
prefragments coincide with the magic numbers, e.g., $N=164$ for
$\alpha_2=0.3$ where $N_1=92$ and $N_2=58$ are both spherical magic
numbers.  Since $N_1+N_2$ is less than the total number of particles
$N$, the rest of the particles which reside in the neck part will be
distributed to the two fragments at the scission, and the fragments
will have the particle numbers larger than $N_j$ supposing that the
prefragment shapes are kept until scission.  This is consistent with
the experimental data for actinide nuclei where the heavier fragments
in most cases have $A\simeq 140$, which is somewhat larger than that of
the doubly-magic $^{132}\mbox{Sn}$.

\begin{table}
\caption{\label{tab:n_frag_a}
Radii $R_j$ of the prefragments in units of $R_0$, and the prefragment
particle numbers $N_j$ relative to the total particle number $N$ for
several values of the mass asymmetry parameter $\alpha_2$, with the
elongation parameter $\sigma_1=2.0$.}
\centering
\tabcolsep=8pt
\begin{tabular}{c|c|c|c|c}
\hline\hline
$\alpha_2$ & $R_1/R_0$ & $N_1/N$ & $R_2/R_0$ & $N_2/N$ \\ \hline
0.1 & 0.78820 & 0.48986 & 0.74973 & 0.42142 \\
0.2 & 0.80607 & 0.52373 & 0.72911 & 0.38760 \\
0.3 & 0.82301 & 0.55747 & 0.70757 & 0.35425 \\
0.4 & 0.83906 & 0.59071 & 0.68509 & 0.32154 \\
0.5 & 0.85420 & 0.62327 & 0.66166 & 0.28967 \\ \hline\hline
\end{tabular}
\end{table}

\begin{figure}
\includegraphics[width=\linewidth]{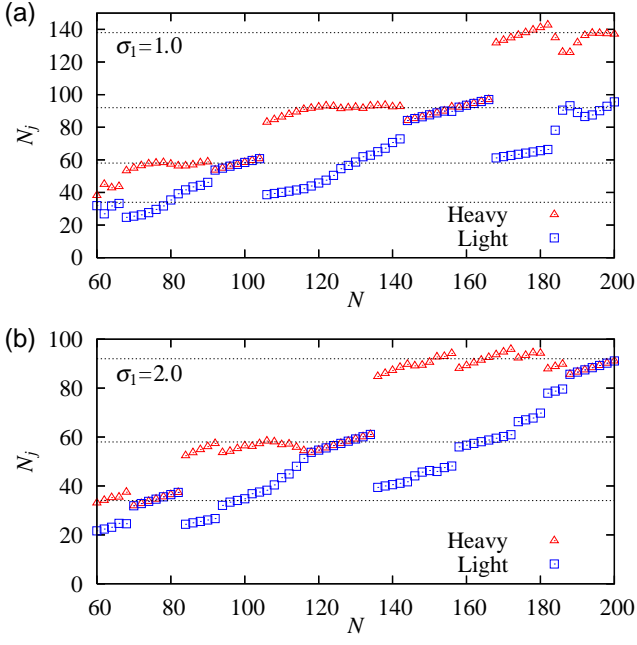}
\caption{\label{fig:abund} 
Prefragment particle numbers $N_1$ (red triangle) and $N_2$ (blue
square) for the mass asymmetry
parameter $\alpha_2$ which minimizes the shell energy for given total
particle number $N$, with the elongation parameter $\sigma_1=1.0$ (a)
and $2.0$ (b).  Horizontal dotted lines indicate the spherical magic
numbers $N=\cdots,34,58,92,138,\cdots$.}
\end{figure}

For a fixed value of the elongation parameter $\sigma_1$, we
calculated the shell energy $\delta E(N)$ as function of the asymmetry
parameter $\alpha_2$ and find the value of $\alpha_2$ which minimizes
the shell energy for each $N$.  Then we evaluate the prefragment
particle numbers $N_j$ by Eq.~(\ref{eq:n_frag}) and plot them as
functions of $N$ in Fig.~\ref{fig:abund}.  Horizontal dotted lines
indicate the spherical magic numbers.  One sees that the particle
number of the heavy fragment sticks to the magic numbers, which causes
asymmetric minima for vast ranges of $N$.  This behavior shows a nice
correspondence with the fragment mass distribution in the fissions of
actinide nuclei.  One will notice that the feature can be found even
for relatively small elongation $\sigma_1=1.0$.  This suggests that
the prefragment shell effect dominates the shell energy already at
rather small elongation where the neck is not sufficiently developed.
It would become effective with the development of neck formation after
competition with the macroscopic energy, and would play an essential
role in determining the shape of the fissioning nucleus.

\subsection{Fission path and the constant-action line}

In a situation where the shell effect is dominated by a certain
periodic-orbit family $\beta$, the shell energy minima will appear
along the constant-action lines (\ref{eq:cac}), which are written as
the constant-length lines (\ref{eq:cac_cavity}) for the cavity system.
As we see in the Fourier analysis, contributions of the prefragment
triangle orbits play a dominant role in deformed shell effect in the
TQS cavity model under consideration.  Therefore, the shell energy
minima are expected to arise along the constant-length lines of the
triangle orbit given by
\begin{equation}
k_F(N)L_{j,31}(\sigma_1,\alpha_2)-\tfrac{\pi}{2}\mu_{31}=2n\pi-\tfrac{\pi}{2}
\end{equation}
with integer $n$; different magic numbers of the prefragments correspond
to the lines with different $n$.
The Fermi level $k_F(N)$ is given by
\begin{gather}
k_F(N)\approx \frac{1}{R_0}\left\{\left(\frac{9\pi N}{4}\right)^{1/3}
+\frac{3\pi B_S}{8}\right\},
\end{gather}
where $B_S$ is the surface area relative to that of the sphere with
the same volume (see Appendix~\ref{sec:app}).

In Fig.~\ref{fig:cac}, contour plots of the shell energies $\delta
E(N)$ are shown for several $N$ as functions of deformation parameters
$\sigma_1$ and $\alpha_2$.  Thick lines represent the constant-length
lines of the prefragment triangle orbits in each of the two
prefragments.  One sees that the shell energy valleys appear along the
constant-length lines and deep energy minima are located around the
crossing points of the lines corresponding to the two prefragments.

For $N=100$ and $164$, one has symmetric minima around $\sigma_1=1$,
that are connected with asymmetric minima around
$(\sigma_1,\alpha_2)=(2,\pm0.3)$ by the constant-length lines.  Those
lines are considered to represent the paths to asymmetric fissions.
In fact, one finds that those lines nicely reproduce the behaviors of
fission paths for actinide nuclei obtained in more realistic
calculations with the TQS parametrization\cite{Ichikawa2012}.  The
above feature is also consistent with what Brack \etal\ obtained in a
different cavity model\cite{Brack97}.

\begin{figure}
\includegraphics[width=.75\linewidth]{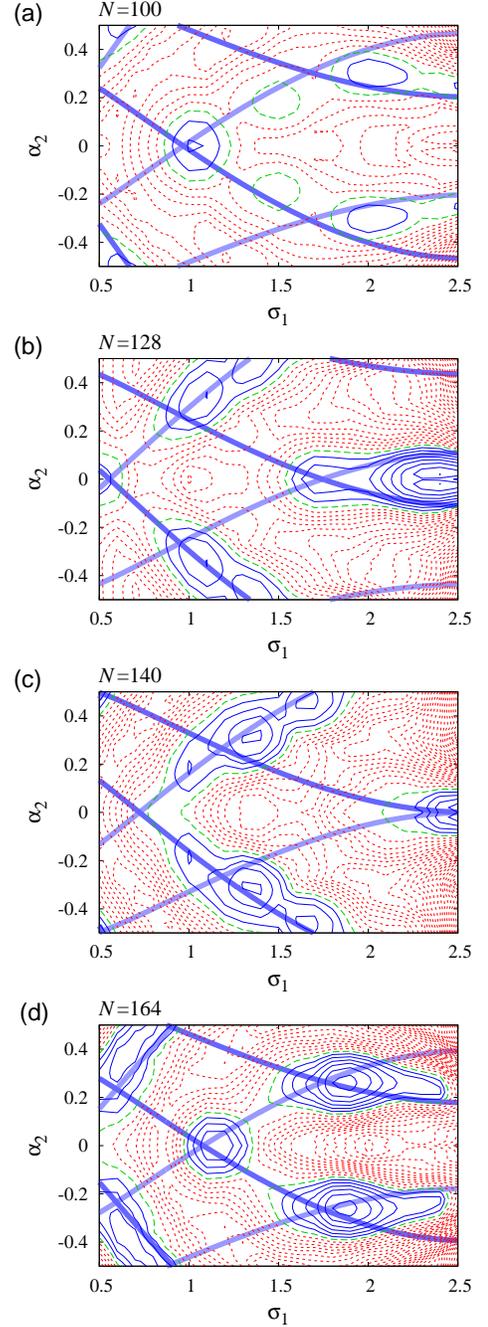}
\caption{\label{fig:cac} 
Contour maps of the shell energy for several particle numbers on the
deformation space $(\sigma_1,\alpha_2)$.  Solid (blue) and dotted
(red) contour lines represent negative and positive shell energies,
respectively.  Thick solid lines represent the constant-action lines
(\ref{eq:cac}) of the prefragment triangle orbits.}
\end{figure}

For $N=128$ and $140$, one has deep energy minima at symmetric
shape with $\sigma_1\gtrsim 2$ but they may not be reached from the
normal-deformed minima around $\sigma_1\approx 0.5$ due to the large
potential barrier.  The shape evolution would be strongly affected by
the potential valley along the constant-length line toward asymmetric
shapes, which leads to the asymmetric fission.

As shown in these calculations, the potential valleys are essentially
determined by the prefragment magicity, and they have significant effect
on the shape of the fissioning nucleus at saddles, that would be
responsible for the fragment mass asymmetry.

\section{Summary}
\label{sec:summary}

Semiclassical origin of the asymmetric fission is investigated with
the idea of prefragment shell effects through the contribution of the
periodic orbits confined in either of the prefragment parts of the
potential.  The cavity potential with the three-quadratic-surfaces (TQS)
shape parametrization is employed to focus attention on the effect of
the shape evolution.  With the use of the trace formula, which was
recently derived for a truncated spherical cavity, we have shown that
the quantum shell energy is nicely reproduced in terms of the
contributions of the prefragment periodic-orbit families.  Taking
notice of the fact that the triangle orbit family makes a dominant
contribution to the shell effect, we have obtained a relation which
expresses the deformed shell energy of the TQS cavity in terms of those of
the isolated spherical fragments.  This relation clarifies the roles
of the prefragment magicity in formation of the fission path in the
potential energy surface.

In a more realistic mean-field potential with finite diffuseness and
spin-orbit coupling, the triangle-type orbit plays a dominant role in
emergence of the distinct spherical magic numbers\cite{Arita2016}.
This suggests that a relation like (\ref{eq:sce_frag_sph}) might be
accessible also in the more general realistic mean-field models.

The prefragment shapes are assumed to be spherical in the present
study, but the effects of the prefragment deformations should be taken
into account in general.  In the realistic macroscopic-microscopic
model analysis using the TQS parametrization, consideration of the
full five dimensional potential energy surface is crucial in obtaining
the correct fission saddles to reproduce the experimental
results\cite{Ichikawa2012}.  (To be more precise, the competitive
effects by macroscopic and shell parts are also found to be
significant in many situations.)  Recently, importance of the octupole
degrees of freedom for the prefragment shape has been
suggested\cite{Scamps2018}.  We expect that the POT analysis is useful
in investigating the stability of the prefragment shape, which is
among the key ingredients in describing the fission dynamics.
Emergence of the remarkable shell structure at a certain deformation
is often related to the local symmetry restoration caused by the
periodic-orbit bifurcation\cite{Sugita97}, around which the amplitude
of the trace formula takes a significant enhancement\cite{Arita2016}.

\appendix
\section{Weyl formula for the average level density}
\label{sec:app}

For a particle with mass $M$ confined in a three-dimensional
infinite-well potential, average level density is given by the Weyl
asymptotic expansion formula\cite{BaBlo2}
\begin{equation}
\bar{g}(e)=\frac{2M}{\hbar^2}\left(\frac{kV}{4\pi^2}
-\frac{S}{16\pi}+\frac{K}{12\pi^2 k}\right).
\end{equation}
Here, $k$ stands for the wave number $k=\sqrt{2Me}/\hbar$, $V$ and $S$
are volume and surface area of the potential well, and $K$ is the
surface integral of the mean curvature
\begin{equation}
K=\oint dS\frac12\left(\frac{1}{R_1}+\frac{1}{R_2}\right),
\end{equation}
with $R_1$ and $R_2$ being the main curvature radii.  For an axially
symmetric surface $\rho=\rho_s(z)$ ($z_{\rm min}\leq z\leq z_{\rm
max}$) with fixed volume $V=4\pi R_0^3/3$, one has
\begin{equation}
\bar{g}(e)=\frac{2MR_0^2}{\hbar^2}\left(\frac{kR_0}{3\pi}
 -\frac{B_S}{4}+\frac{B_K}{3\pi kR_0}\right)
\label{eq:sld_av}
\end{equation}
where $B_S$ and $B_K$ are the surface area and the surface integral of
the mean curvature relative to those for spherical shape,
\begin{gather}
B_S=\frac{1}{2R_0^2}\int_{z_{\rm min}}^{z_{\rm max}}
 \rho_s\sqrt{1+(d\rho_s/dz)^2}dz \\
B_K=\frac{1}{4R_0}\int_{z_{\rm min}}^{z_{\rm max}}
 \left\{1-\frac{\rho_s(d^2\rho_s/dz^2)}{1+(d\rho_s/dz)^2}\right\}dz
\end{gather}
Using the wave number variable $k$, one has
\begin{equation}
\bar{g}(k)=2R_0\left\{\frac{(kR_0)^2}{3\pi}
-\frac{B_SkR_0}{4}+\frac{B_K}{3\pi}\right\}.
\end{equation}
The average particle number filled up to the Fermi level $k=k_F$,
taking account of the spin degeneracy factor, is given by
\begin{align}
&\bar{N}(k_F)=2\int_0^{k_F}\bar{g}(k)dk \nonumber \\
&\quad =\frac{4(k_FR_0)^3}{9\pi}
-\frac{B_S(k_FR_0)^2}{2}+\frac{4B_K k_FR_0}{3\pi},
\end{align}
which is inverted as
\begin{equation}
k_F(N)\approx \frac{1}{R_0}\left\{
\left(\frac{9\pi N}{4}\right)^{1/3}+\frac{3\pi B_S}{8}\right\}
+\mathcal{O}(N^{-1/3}).
\end{equation}

\end{document}